# Purcell enhanced emission and saturable absorption of cavity-coupled CsPbBr$_3$ quantum dots

Purbita Purkayastha [1,2], Shaun Gallagher [4], Yuxi Jiang [1,3], Chang-Min Lee[1,3], Gillian Shen [4,5], David Ginger [4], Edo Waks [1,3]

[1]Institute for Research in Electronics and Applied Physics and Joint Quantum Institute, University of Maryland, College Park, Maryland 20742, USA
[2]Department of Physics, University of Maryland, College Park, Maryland 20742, USA
[3]Department of Electrical and Computer Engineering, University of Maryland, College Park, Maryland 20740, USA
[4] Department of Chemistry, University of Washington, Seattle, Washington 98195, USA
[5] Department of Materials Science and Engineering, University of Washington, Seattle, Washington 98195, USA

## Abstract :

Halide perovskite semiconductors have emerged as promising materials for the development of solution-processed, scalable, high performance optoelectronic devices such as light-emitting diodes (LEDs) as well as coherent single photon emitters. Their integration to nanophotonic cavities for radiative enhancement and strong nonlinearity is underexplored. In this work, we demonstrate cavity-enhanced emission and saturable absorption using colloidal CsPbBr$_3$ perovskite quantum dots coupled to a high-Q cavity mode of a circular Bragg grating structure designed to facilitate integration of solution-processed materials . We achieve an order of magnitude increase in brightness and 8-fold increase in the spontaneous emission rate for the cavity-coupled emitters. This result indicates the possibility of achieving transform-limited photon coherence for the halide perovskites at cryogenic temperatures. We also observe saturable absorption of the emitters through intensity-dependent cavity quality factor. These results pave the way towards achieving improved photon indistinguishability and strong optical nonlinearities for cavity coupled perovskite systems.

**Keywords:** Halide perovskites, circular Bragg Gratings, photonic crystals cavities, nanophotonics, colloidal quantum dots, optical nonlinearity.

## Introduction:

Halide perovskite quantum dots are promising light emitters because of their bright emission, color tunable optical properties, low-cost chemical synthesis,[1] and low blinking and spectral diffusion.[2] They can also be easily integrated to a variety of nanophotonics platforms to improve their emission properties.[3] In addition, halide perovskites show significant promise for the development of high-performance photonic and energy devices that rely on optical nonlinearity.[4,5] They have also been shown to be efficient single photon sources with relatively long coherence time of 80 ps,[6] very close to the radiative lifetime limit of 150 − 300 ps at cryogenic temperatures.[6–8] They are therefore excellent candidates for generation of indistinguishable single photons.

The integration of halide perovskites with nanophotonic cavities can enhance optical emission and enable strong optical nonlinearities. Nanophotonic cavities can improve the spontaneous emission rate through the Purcell effect [9] [10] and can also direct emission more efficiently out of the surface.[11] These properties combine to enhance emitter brightness. [12,13] Although a number of studies have reported coupling of perovskite quantum dots to cavities, the Purcell factor, reflecting the relative increase of the spontaneous emission rate in the cavity relative to free space, for these studies have generally been below 3.[14–17] The coupling of halide perovskites to cavities can also generate strong optical nonlinearity through saturable absorption. Previous works have studied saturable absorption in perovskite films,[18,19] but the integration of this nonlinearity into optical cavities remains unexplored.

In this letter, we report enhanced spontaneous emission and saturable absorption from halide perovskites coupled to a nanophotonic cavity. We integrate $CsPbBr_3$ quantum dots capped with zwitterionic ligands into a circular bullseye cavity with a mode exhibiting a quality factor of 1980. We achieve an order of magnitude improvement in the brightness of the cavity coupled dots compared to dots on an unpatterned surface. From time-resolved lifetime measurement, we observe an average cavity-enhanced lifetime of 20 ps, which gives us a Purcell factor of 8. This lifetime is below the 80 ps dephasing time previously reported,[6] suggesting a pathway towards indistinguishable single photon emission. We also observe saturable absorption through an intensity-dependent cavity quality factor. These findings are an important step towards utilizing cavity coupled perovskites as efficient light sources and for strong optical nonlinearity.

We synthesized the $CsPbBr_3$ quantum dots used in this work using a room-temperature synthesis employing trioctylphosphine oxide as a capping ligand as reported in previous work.[20] This synthesis produces a narrow size distribution of spheroidal quantum dots by decoupling nucleation and growth.[21] Following the initial synthesis we performed a surface ligand exchange with lecithin, a zwitterionic ligand previously shown to stabilize colloidal $CsPbBr_3$ nanoparticles,[22] as described in the SI.

Fig1a shows the TEM image of the synthesized quantum dots. As expected based on the synthetic route, these quantum dots have a spheroidal shape,[20] with a size distribution of $9.6 \pm 1.0$ $nm$. Fig1b shows ensemble absorption and emission spectra for the perovskite quantum dots in toluene. The quantum dots exhibit bright photoluminescence with peak emission at 509 nm and a full-width at half-maximum linewidth of 16 nm at room temperature. The photoluminescence quantum yield for this batch of quantum dots was 94% measured in solution. We observe a ~$10 - 15$ nm redshift of their emission peak when cooled from room T to 3.2K, due to a phase transition commonly observed in lead halide perovskites.[23] At the same time, the ensemble linewidth narrows to around 4 nm, which is dominated by inhomogeneous broadening.

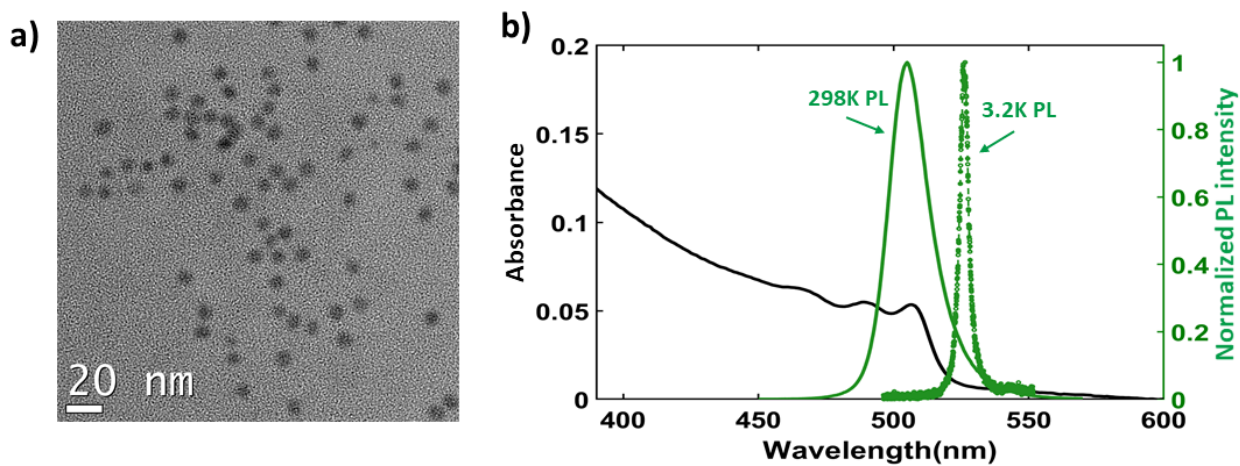

**Figure1**: **a)** TEM image of perovskite quantum dots. **b)** Absorbance (black) and photoluminescence(green) spectra of perovskites dispersed in toluene. The solid green curve is the photoluminescence at room temperature in toluene solution, and the dashed green curve is the photoluminescence at 3.2K.

To enhance the emission of the CsPbBr$_3$ quantum dots, we fabricated planar supported bullseye structures as illustrated Fig.2a. This planar cavity geometry with a supported bullseye enables facile incorporation of solution-processed emitters like colloidal quantum dots following fabrication of the cavity. We patterned these cavities in SiN, which is transparent at the emission wavelength of the emitter. The bullseye structure consists of a central disc of diameter '$D$', with etched concentric rings of periodicity '$\Lambda$' and gap width '$\Delta$'. The cavity features six symmetrically placed unetched "bridges" that are tapered from the central disc at an angle '$\theta$'. These bridges mechanically support the cavity structure. The periodic rings serve as Bragg mirrors, restricting light confinement in the in-plane direction.[24,25]

We optimized the design parameters numerically using Finite Difference Time Domain (FDTD) simulations to attain a cavity resonance which matches the perovskite emission at low temperature. We determined the optimal cavity parameters to be $D = 440$ nm, $\Lambda = 180$ and $\Delta = 84$ nm for a 120 nm thick SiN membrane. We fixed the bridge angle $\theta$ at 3 degrees, which is sufficient to mechanically suspend the membrane without significantly degrading the cavity quality factor ($Q$). With these parameters, we observed that the cavity supports two modes, with the lowest order mode at 530 nm and a higher order mode at 511 nm. The first order mode at 530 nm has a mode profile as shown in Fig 2b and directional far field emission (inset of Fig 2b). But the simulated Q factor of this mode is low (around 400). On the other hand, the second order mode at 511 nm has a much higher Q of 3500 but less directionality. Fig 2c shows the mode profile and the farfield emission profile of the second order mode. Although the $2^{nd}$ order mode is less directional, we can still collect ~23% of emission falling inside 45 degrees defined by 0.7 numerical aperture of our objective lens. Hence, we determined that the $2^{nd}$ order mode appears to be a better choice to achieve a high Purcell effect.

We fabricated the cavity using e-beam lithography and inductively coupled plasma etching. First, we deposited a 120nm layer of SiN by stoichiometric low pressure chemical vapor deposition (LPCVD) on a substrate of Si with a 2 $\mu m$ SiO$_2$ on top. We then patterned the bullseye design on SiN with standard e-beam lithography using a ZEP 520 mask, followed by Fluorine-based inductively coupled plasma reactive dry etching using a CHF$_3$ / SF$_6$ gas mixture. We undercut the etched structure by wet-etching the sacrificial SiO$_2$ layer using buffered oxide solution. Fig 2d shows a scanning electron microscope image of the resulting suspended cavity. After fabricating the cavity, we deposit the CsPbBr$_3$ perovskites by drop casting a toluene and 2% polystyrene(by weight) solution with dot concentration of ~0.26 mg/ml w/v onto the cavity surface.

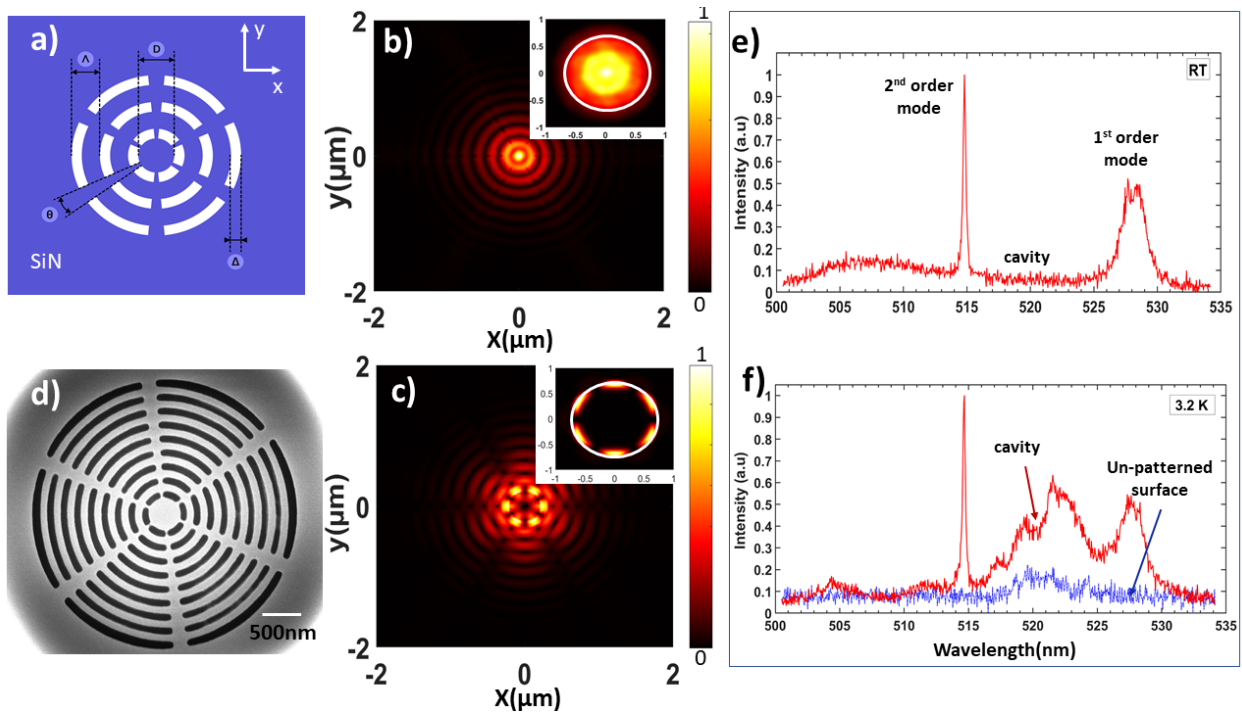

**Figure 2**. **a)** Schematic diagram of bullseye optical cavity on SiN . **b,c)** Finite difference time domain (FDTD) simulation of first order mode profile at 530 nm and 2nd order mode profile at 511 nm . Insets show the respective farfield profile and the white circle in the insets indicate the N.A=0.7 of collection objective lens. The color map indicates the normalized electric field intensity value **d)** Scanning electron microscope image of the fabricated device. **e)** Room temperature characterization of the cavity modes. **f)** Cavity enhanced emission of the perovskites(red) and perovskite emission from unpatterned surface (blue) at 3.2K .

We performed all the measurements in a confocal microscope setup that can operate at both room temperature and $3.2\ K$. We excited the sample above the bandgap using a continuous wave laser at 405 nm and collected the signal from the sample using an objective lens with a numerical aperture of $0.7$. We filtered the collected emission signal with a dichroic mirror of cutoff at 490 nm and further eliminated the laser with a notch filter at 405 nm. We measured the photoluminescence using a grating spectrometer with a spectral resolution of $0.02$ nm.

Figure 2e and 2f show the photoluminescence spectra from a fabricated cavity at both room-temperature and $3.2\ K$. Figure 2e shows the photoluminescence spectrum of the cavity at room temperature when excited with high laser power (~120µW/µ$m^2$). We observed both the low $Q$ first order mode and higher $Q$ second order mode, both of which are labeled in the figure. By fitting to a Lorentzian function, we calculate the quality factor of the lowest order mode to be $Q = 240$, and the higher order mode to be $Q = 1980 \pm 114$. The red data points in Figure 2f show the photoluminescence from the quantum dots in the same cavity at $3.2\ K$ with about 12.7 µW/µ$m^2$ excitation power. For comparison we also plot the photoluminescence emission from an unpatterned region of the device (blue circles). The emission from the higher order cavity mode is an order of magnitude brighter than the unpatterned region. When averaged over multiple cavities, we observe about 9 times improvement in the emission intensity.

To characterize the enhancement of spontaneous emission, we next conducted time-resolved photoluminescence lifetime measurements. We performed the photoluminescence lifetime measurements using excitation pulses at 405 nm generated by frequency doubling a Ti: Sapphire laser. The pulsed excitation repetition rate was 76 MHz and pulse duration of 2 ps. We conducted the measurements below saturation and in the regime where emission intensity is linearly proportional to excitation power. We filtered the emission through a spectrometer grating and slit and the signal was detected using an avalanche photodiode(APD). Subsequently, we used a time correlated single photon counting setup (TCSPC) (Picoharp 300) for signal measurement.

Figure 3a compares the time resolved lifetime measurement on a cavity and a region on the unpatterned surface at $3.2\ K$. The APD used for measurement has a temporal resolution of better than 50 ps. To attain better temporal resolution, we conducted reconvolution of lifetime response with detector response using Easytau software(Instrument Response Function correction). We normalized the decay curves with respect to counts at t=0.

We fitted the decay from the cavity and unpatterned surface to a stretched exponential function, as it can provide a good description of average lifetimes from a heterogenous population of emitters.[26][27][28]

$$I(t) = I_0 + A\ exp(-t/\tau)^\beta \quad (1)$$

Here, $I(t)$ represents the photoluminescence intensity at time t, $I_0$ is the background intensity, $A$ is the emission intensity at time zero, τ is the characteristic lifetime, and β is the stretching parameter reflecting heterogeneity in the lifetime of the emitter population, varying within the range of 0 to 1. The stretched exponential model describes an ensemble of emitters with different lifetimes. It is thus highly suitable for quantum dots coupled to the cavity mode, since each dot exhibits a different Purcell effect due to variations in the spatial position and dipole orientations.[26][29] From the fit, we determined $\tau = 25.2$ +/- $0.8$ ps and $\beta = 0.9$ for the perovskites coupled to cavity and $\tau = 132.4$ +/- $6.4$ ps and $\beta = 0.5$ for the perovskite quantum dots on the unpatterned SiN surface. Using these parameters, we can calculate the average lifetime as $T_{avg} = \tau/\beta * \Gamma(1/\beta)$ where $\Gamma(z)$ is the gamma function. From this equation, we determine the average lifetime of the cavity and unpatterned surface to be 26 ps and 265 ps respectively. These results indicate a lifetime enhancement of 10 for this particular cavity.

To get a more accurate statistical estimate of the Purcell enhancement, we performed measurements on multiple cavities and unpatterned regions. Figure 3b plots a histogram of lifetimes for 12 different cavity structures and unpatterned surface regions at $3.2 K$. All the cavity structures exhibit significantly shorter lifetimes than all of the unpatterned regions. From the average of the two groups, we find that the average cavity lifetime is 20.9 +/- 9.39 ps, while the average lifetime of the unpatterned surface is 168.03 +/- 49 ps. These values correspond to an average Purcell factor of 8. The radiative lifetimes of all our measured cavities are below 80 ps, which is the coherence time previously measured for these emitters.[6] These values suggest that the cavities are able to achieve lifetime-limited photon emission required to generate indistinguishable single photons. The average lifetime of our dots on the unpatterned surface (168.03 +/- 49 ps) is consistent but on the lower end of the range of previously reported values of ($\sim 150 - 300$ ps [6–8]). This slight reduction could be consistent with recent theory work that has predicted that spheroidal dots should exhibit shorter lifetimes compared with their cubic counterparts due to difference in dielectric confinement of the photon field in a sphere compared to a cube.[30]

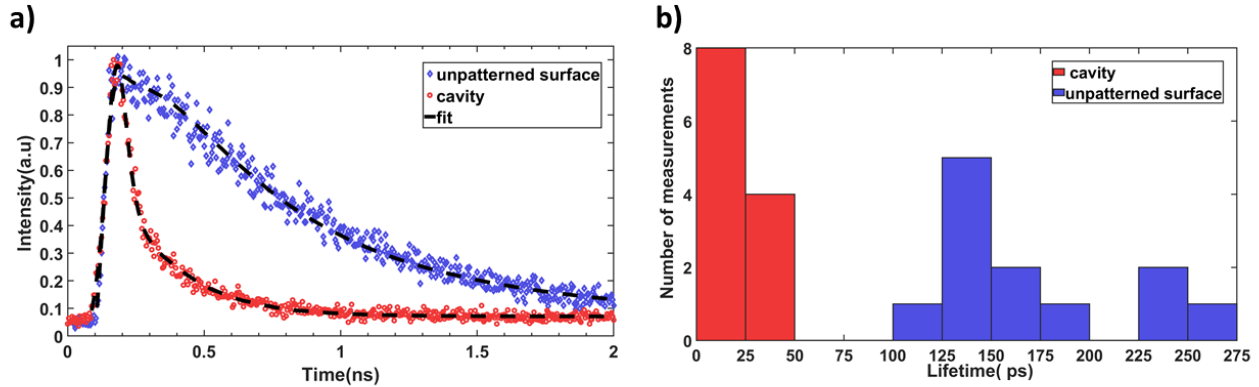

**Figure 3: a)** Time-resolved photoluminescence decay of cavity coupled emitter (blue) under above band pulsed excitation and a comparison measurement for the emitters on the unpatterned region(red). The black dashed lines give the stretched exponential fit of the decay curves after instrument response correction. **b)** Histogram of the lifetimes after curve fitting from 12 different cavities and 12 bulk regions.

In addition to the Purcell effect, the cavities should also exhibit strong optical nonlinearity due to saturable absorption.[31] This is because, the saturable absorbers can concentrate transmitted light in the cavity non-linearly, which modifies the optical losses of a cavity resulting in change in cavity quality factor. This saturable absorption has a number of important applications including optical switching [32,33], optical bistability,[34] and mode locking [35,36] etc. To demonstrate saturable absorption we increased the concentration of quantum dots by an order of magnitude(~2.6mg/ml quantum dots in toluene and 2% polystyrene). The quantum dots were deposited using the same drop-casting technique used for Purcell effect measurements. We performed all saturable absorption measurements at room temperature by exciting the cavity using $405 nm$ light from a pulsed laser with 76 MHz repetition rate and collecting the photoluminescence through a grating spectrometer with spectral resolution of 0.02nm. The laser excitation spot had an estimated diameter of 1 μm.

Figure 4a plots the integrated intensity of the cavity emission as a function of peak pump power. We measured the total intensity by fitting the cavity enhanced emission peak to a Lorentzian and calculating the area under it. The solid curve is a fit to a numerical model for saturation, given by -

$$I = I_0(1 - exp(-P/P_{sat}))  \quad (2)$$

From the fit, we obtain the saturation intensity to be $P_{sat} = 75.7 mW$. This corresponds to a saturation fluence of ~ 19.2 $\mu J/cm^2/pulse$. The intensity shows a clear saturation behavior, indicating that the device is operating below lasing threshold.

Figure 4b plots the cavity quality factor as a function of peak pump power. We measure the cavity $Q$ by fitting the cavity peak to a lorentzian and calculating the linewidth and peak wavelength $(Q = \lambda/\Delta\lambda)$. The $Q$ increases from an initial value of 980 at low power to its final saturated value of 1200 at a peak power of 800 mW. This value represents a 20% increase in the quality factor. This change in cavity $Q$ is caused by saturable absorption of emitters in the cavity, which is expected when the emitters are below lasing threshold. Because we are below the lasing threshold, the increase in quality factor is not due to Schawlow-Townes linewidth narrowing[37].

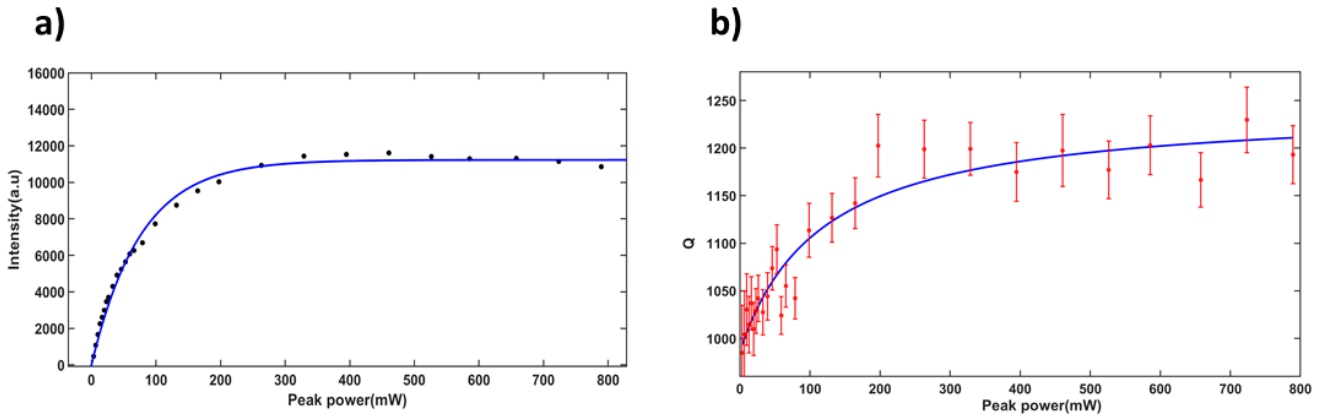

**Figure 4.** a) Integrated cavity photoluminescence intensity as a function of input pump power. The blue line is a fit using the model equation -2 b) Cavity Q as a function of input pump power. The blue line is a fit using the model equation described in equation-3.

To quantify the non-linearity of the cavity, we fit the change in cavity Q with peak pump power to a model equation based on cavity coupled to saturable absorber [38]-

$$\frac{1}{Q} = \frac{1}{Q_c} + \frac{1/Q_{ab}}{1+P/P_{sat}} \quad (3)$$

where $Q_c$ is the bare cavity quality factor, $Q_{ab}$ is the quality factor due to quantum dot absorption, $P$ and $P_{sat}$ are the peak input and saturation powers respectively. From the fit, we measure a saturation power of $P_{sat} = 91.7 mW$ (corresponding to a fluence of ~ $23.3 \mu J/cm^2/pulse$), $Q_{ab} = 4760$ and $Q_c = 1244$. The slight difference in the saturation fluence for cavity quality and quantum dot saturation might arise because of the errors introduced while measuring the linewidths of the emission peak for quantifying cavity quality. In another study that incorporated a different colloidal saturable absorber non-linearity into optical cavity investigations, comparable saturation fluence was observed. However, the quantum dots utilized in that research proved to be excessively unstable, rendering them unsuitable for practical applications.[39] We note that the quality factors of the cavities measured here with a high quantum dot concentration are smaller compared to the quality factors of the cavities with lower dot densities that were used to perform Purcell enhancement measurements in this work. We attribute this difference to the fact that higher dot densities may introduce additional undesirable loss due to dielectric scattering.

## Conclusion:

In conclusion, we demonstrate cavity enhanced photon emission and saturable absorption for an ensemble of $CsPbBr_3$ quantum dots. We achieve an order of magnitude enhancement in intensity and an average Purcell enhancement factor of 8, potentially improving the photon indistinguishability for the emitters coupled to a relatively high Q mode of circular bullseye cavity. We also measured saturable absorption of the dots through cavity quality where the cavity Q increases by around 20%. The nonlinearity in this device may find use in a variety of applications including all-optical switching, [33,40] pulsed laser generation,[41] and optical memories.[42] Increasing both the cavity quality factor and the emitter concentration could enable the device to achieve lasing, implementing a compact solution-processed on-chip coherent light source. Ultimately, this work represents important progress towards utilizing halide perovskite quantum dots as coherent photon sources and low power nonlinearities for photonic integration.

## Acknowledgements:

This material is based primarily upon work supported by the National Science Foundation under Grant No. DMR-2019444, including the roles of P.P., E.W., S.G. G.S, and D.S.G. We would like to thank Tom Loughran for help with SiN deposition and Jessica Kline for discussions on the results.

The UMD authors acknowledge additional funding support from the AFOSR grant #FA95502010250 and #FA95502310264 which supported Y.J., and the Maryland-ARL quantum partnership #W911NF1920181 which supported E.W.

The UW authors also acknowledge the use of facilities and instruments at the Photonics Research Center (PRC) at the Department of Chemistry, University of Washington, as well as that at the Research Training Testbed (RTT), part of the Washington Clean Energy Testbeds system supported by the state funded Clean Energy Institute.

## Competing financial interests:

The authors declare no competing financial interests.

# Methods:

**Ensemble QD Characterization:**

Absorbance spectra of the QD solutions were collected with a Perkin-Elmer Lambda 950 UV/Vis/NIR Spectrometer in a range of 200-1100 nm with an integration time of 0.5 s.
Steady-state photoluminescence (PL) spectra were acquired using Perkin-Elmer Fluorescence Spectrometer LS 55. Spectra were collected using an excitation wavelength of 405 nm. The emission bandwidth was kept at 3 nm and dwell time at 0.2 s.
Photoluminescence quantum yield measurements (PLQY) were performed on a commercial integrating sphere system (Hamamatsu Photonics K.K). PLQY values are determined using a white light source (Hamamatsu Mercury Xenon Lamp) and a monochromator for wavelength selection (405 nm) as the excitation source to illuminate the samples in an integrating sphere (Hamamatsu Photonics K.K). The optical density of samples was kept below 0.05 at the excitation wavelength (405nm) to minimize reabsorption effects. Spectral correction was performed using a commercial white light source (Ocean Insight HL-3P-INT-CAL) to correct for the responsivity of the detector. PLQY was calculated using the following formula:

$$PLQY = \frac{I_{em,\ sample} - I_{em,\ blank}}{I_{ex,\ blank} - I_{ex,\ sample}} * 100 \qquad (4)$$

Where $I_{em,sample}$ and $I_{em,blank}$ are the integrated area under the curve in the emission region (450-600 nm) of the sample and the neat hexane blank, respectively. The $I_{ex,sample}$ and $I_{ex,blank}$ are the integrated area under the curve in the excitation region (395-415 nm) of the sample and the hexane blank respectively.

**Device Fabrication:** To fabricate the Bullseye design, we first deposited a 120 nm thick SiN membrane on a Si wafer with 2 micron $SiO_2$ using stoichiometric Low Pressure Chemical Vapour Deposition(LPCVD) at a rate of 0.4 nm/min. We then spin coated a thin layer of positive electron beam resist(ZEP- 520A) at a rate of 2600 rpm for 1 min and post baked at 180 degree celsius for 3 mins. We patterned the Bullseye design on the resist using e-beam lithography( Elionix ELS- G100) . Then, we developed the structure by cold-develop technique using xylene as developer for 2.5 mins inside a freezer followed by baking at 120 degree celsius for 2 mins. After developing, we used inductively coupled Fluorine-based plasma dry etching with CH3/SF6 gas mixture to etch the nitride layer. The etch rate for the process is 2.7 nm/min. After etching the nitride , we used a wet etching recipe using 6:1 buffered oxide(BOE) to remove the sacrificial SiO2 layer and make the structure suspended. On the fabricated device, we drop-casted a solution of $CsPbBr_3$ quantum dots suspended in hexane and diluted in toluene and 5% polystyrene. We let the solution dry in air for a few minutes before dabbing the edges for the solvent to dry completely.

**Experimental Setup:** The samples were mounted inside a Montana cryostat to characterize the cavities and perform the measurements both under vacuum at room temperature as well as at cryogenic temperature( 3.2K). The objective lens used in the setup has a numerical aperture of 0.7 and focal length of 4mm. The samples were excited with an above-band continuous wave laser diode(from Thorlabs) at 405 nm as well as with a Ti:Sapphire MIRA laser operating at 810 nm and frequency doubled with a BBO crystal. The collected emission spectra was spectrally measured using a spectrometer (Princeton Instrumentation) with grating density 1800g/nm and 600g/nm, and a 1340 pixels CCD detector providing a spectral resolution of 0.02nm. We performed the lifetime measurements using the spectrometer gratings as filters and collecting the photons into an Avalanche Photo diode ( Micro Photon Devices). We used a time interval analyzer (Picoharp 300) to record the intensity versus time. To measure the instrument response function (IRF) of the detector components, short pulses from the MIRA laser close to the emission wavelength were recorded in Picoharp . Then, we performed exponential deconvolution operation on the recorded lifetimes with IRF using the EasyTau software, for obtaining better temporal resolution.